\UseRawInputEncoding
\documentclass[
reprint, 
showpacs,
nobibnotes,
amsmath,amssymb,
aps, 
pra,
]{revtex4-2}

\usepackage{graphicx}
\usepackage{dcolumn}
\usepackage{bm}
\usepackage{feyn}
\usepackage{subfigure}
\usepackage{natbib}
\usepackage{float}
\usepackage{color}
\usepackage{hyperref}
\usepackage{mathrsfs}
\usepackage{rotating}
\usepackage{epsfig}
\newcommand{\RNum}[1]{\uppercase\expandafter{\romannumeral #1\relax}}
\newcommand{\be}{\begin{eqnarray}}
	\newcommand{\ee}{\end{eqnarray}}

\begin{document}



\title{Imaginary coupling induced Dirac points and group velocity control in non-reciprocal Hermitian Lattice}

\author{Yuandan Wang}
\affiliation{School of Physics, Northwest University, 710127, Xi'an, China}
\author{Junhao Yang}
\affiliation{School of Physics, Northwest University, 710127, Xi'an, China}
\author{Yu Dang}
\affiliation{School of Physics, Northwest University, 710127, Xi'an, China}
\author{Haohao Wang}
\affiliation{School of Physics, Northwest University, 710127, Xi'an, China}
\author{Guoguo Xin}
\affiliation{School of Physics, Northwest University, 710127, Xi'an, China}
\author{Xinyuan Qi}
\email{ qixycn@nwu.edu.cn}
\affiliation{School of Physics, Northwest University, 710127, Xi'an, China}

\date{\today}

\date{\today}
\begin{abstract}
We propose a mechanism to achieve the group velocity control of bifurcation light via an imaginary coupling effect in the non-reciprocal lattice. The physical model is composed of two-layer photonic lattices with non-reciprocal coupling in each unit cell, which can support a real energy spectrum with a pair of Dirac points in the first Brillouin zone due to the Hermicity. Furthermore, we show that the systems experience topological phase transition at the Dirac points by tuning the coupling strength, allowing the existence of topological edge states on the left or right boundaries of respective lattice layers. By adjusting the imaginary coupling and the wave number, the group velocity of the light wave can be manipulated, and bifurcation light transmission can be achieved both at the Dirac points and the condition without the group velocity dispersion. Our work might guide the design of photonic directional couplers with group velocity control functions.
\end{abstract}


\maketitle


\section{Introduction}
Over the past few decades, group velocity adjusting has been the focus of attention as an important technique in optical communications, nonlinear optics, optical storage, et.al~\cite{stenner2005fast,dumeige2011quasi,iliew2008slow,su2011dynamics,krauss2008we}. In traditional Hermitian systems, atomic systems based on electromagnetically induced transparency effects are used to generate fast and slow light~\cite{1999Light,1999Ultraslow,bajcsy2003stationary,everett2017dynamical}, which can enhance the sensitivity of interferometers and sensors~\cite{scheuer2006sagnac,baker2012slow, Qin2016Slow,shahriar2007ultrahigh}. In the photonic system, photonic crystals~\cite{yan2017slow,zhang2004phase,baba2008slow,2022High}, photonic lattices~\cite{longhi2015robust,GARANOVICH20121,li2022anisotropic}, and optical fibers~\cite{2005Tunable,thevenaz2008slow,2011Superluminal}  can all realize the control of the group velocity of light pulse. Among them, photonic lattice, an important platform for adjusting the group velocity, has been studied intensively due to its compactness and multi-functionality\cite{fleischer2003observation,mukherjee2018experimental}. Periodic oscillation of the group velocity and dynamical localization of the light wave can be achieved by introducing the perturbation to one-dimensional photonic lattices~\cite{2019Hermiticity}. Considering the direction of coupling effects, stopping surface magneto-plasmons at terahertz frequencies is realized when the coupling coefficients in the two-dimensional photonic lattice are non-reciprocal~\cite{2021Stopping}. However, these systems are non-Hermitian once they are perturbated or the coupling is non-reciprical~\cite{budich2020non,zhang2022universal,zhang2021superior}. Therefore, constructing a non-reciprocal Hermitian system to achieve group velocity regulation will be a worthy topic.

Dirac Points (DPs)~\cite{Xue2020Non,liu2020observation,ozawa2019topological}, also known as Devil Points, are an emerging research hotspot in the field of optics in recent years~\cite{he2015emergence,xie2014trapped,guo2017three}.  DP is expressed as the intersection of two energy bands that linearly intersect to form a Dirac cone in the Brillouin zone, which also might be the phase transition point of topological materials~\cite{lu2014topological, ozawa2019topological, qi2011topological}. Current research on DPs focuses on two-dimensional systems. Generally, four types of DPs are investigated. (1) High-symmetry DPs, as the first type, contribute to extraordinary transport of electrons or photons, such as Klein tunneling and weak antilocalization arising from the linear crossings in their band structure~\cite{young2015dirac, he2015emergence, neto2009electronic,2015Creating, 2019Type}. (2) Asymmetric DP, obtained by introducing time-reversal symmetry breaking at the Brillouin zone boundaries, can result in valley-induced unidirectional electromagnetic propagations~\cite{Haldane1988Model, Leykam2016Anomalous, liu2020observation}. (3) Movable DPs, located at the generic momentum points, are movable in the momentum space, e.g., by adjusting the lattice strains~\cite{Lu2022Realization, Lu2016Multiple}. (4) Dirac-like points, such as the intersection of the triple degenerate Dirac-like cone, can be used to realize the constant group velocity propagation of light pulse~\cite{huang2014sufficient, Xu2020Pulse,huang2011dirac,yang2016experimental}. Only a few works on DPs in one-dimensional or quasi-one-dimensional photonic systems are investigated, such as the diffractionless self-splitting light 
wave transmission due to the unique dispersion relationship near the DPs~\cite{zeuner2012optical,li2022subspace}. Even so, the group velocity adjustment mechanism of these bifurcated lights has not been reported yet.

In this work, we propose a method to realize Hermitian systems using non-reciprocal coupling. We obtain a real energy spectrum with a pair of tunable DPs by introducing an imaginary coupling coefficient inside each cell. By adjusting the coupling coefficient, the position and spacing of the double DPs in the energy spectrum can be changed, and finally, these two points will degenerate into one. Correspondingly, the topological states of the system could also be controlled. The bifurcated diffractionfree light transmission with different group velocity distributions can be achieved both at the DPs and without the group velocity dispersion.

\section{THEORETICAL MODEL}

\begin{figure}[htp]
	\centering
	\includegraphics[width=\linewidth]{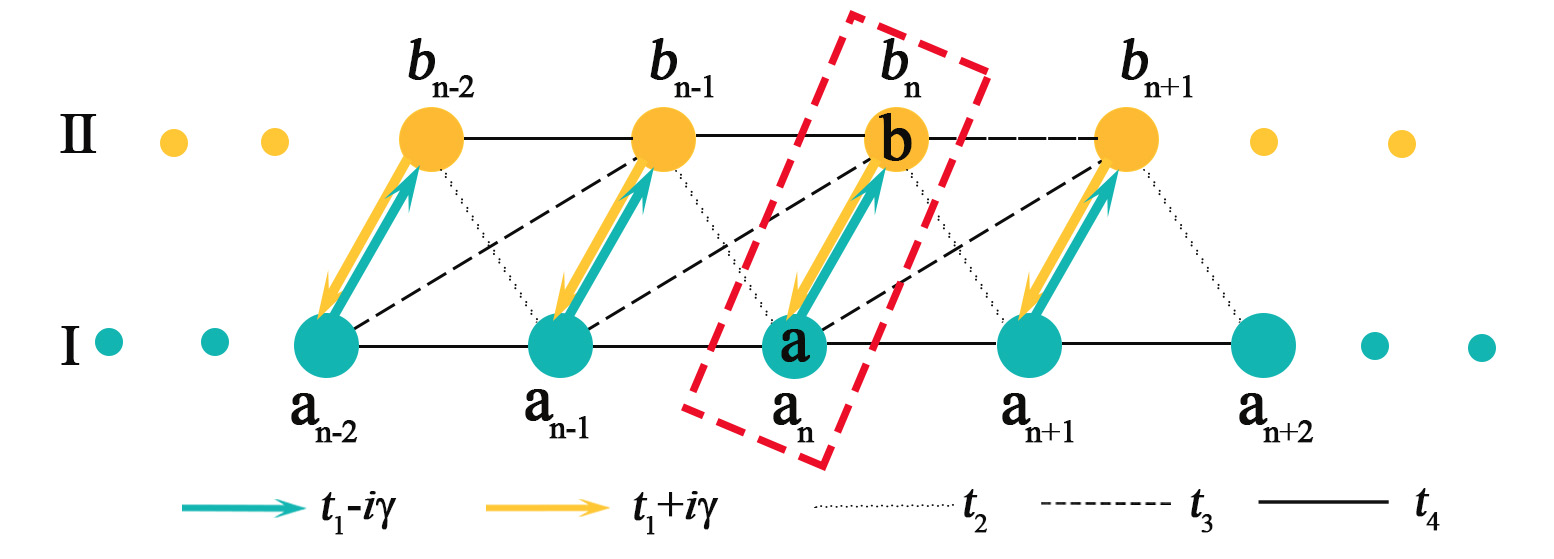}
	\caption{Schematic of the quasi-one-dimensional Hermitian photonic lattice with non-reciprocal coupling efficient. The red dotted box represents a unit cell. }
	\label{fig.1}
\end{figure}
A two-layered photonic lattice composed of an array of periodically arranged evanescently coupled waveguides is shown in Fig. 1, which consists of two sites ($a_n$ and $b_n$) per unit cell. In coupled-mode theory, our single mode-coupled waveguide lattice can be modeled by a tight-binding system as follows:
\begin{align}
	\label{eq:1}
	-i \frac{d a_n}{d z}&=(t_1+i\gamma)b_n+t_4(a_{n+1}+a_{n-1})\notag\\
	&+t_3b_{n+1}+t_2b_{n-1}, \\
	\label{eq:2}
 	-i \frac{d b_n}{d z}&=(t_1-i\gamma)a_n+t_4(b_{n+1}+b_{n-1})\notag\\
 	&+t_3a_{n-1}+t_2a_{n+1},
\end{align}
where $a_n$ and $b_n$ represent the field amplitudes of the first- and second-layer waveguides, respectively. $n$ is the cell number. $t_1$, $t_2$ and $t_3$ are the interlayer coupling constants and $t_4$ is the intralayer coupling, $\pm i\gamma$ are the imaginary coupling coefficient. 

This arrangement can be described by the discrete Schr\"{o}dinger equation $i(d/dz)\psi_n = H_k\psi_n$, where $z$ denotes the propagation coordinate and $\psi_n = (a_n, b_n)^T$ is the two-component wave function describing the field amplitude in each unit cell. The momentum space Hamiltonian $\boldsymbol{H(k)}$ is given by
\begin{align}
	\label{eq:3}     \boldsymbol{H(k)}&=d_0\sigma_0+d_x\sigma_x+d_y\sigma_y,\\
	\label{eq:4} 	d_0(k)&=2t_4\cos k,\\
	\label{eq:5}    d_x(k)&=t_1+(t_2+t_3)\cos k,\\
	\label{eq:6} 	d_y(k)&=(t_2-t_3)\sin k-\gamma,
\end{align}
where  $k$ is the Bloch wave number in the first Brillouin zone. $\sigma_0$ is the identity matrix, $\sigma_{x,y}$ are the Pauli matrices. Although the intracell coupling coefficient $t_1\pm i\gamma$ is non-reciprocal, the system is still Hermitian, satisfying $H=H^{\dagger}$. Notice that the diagonal elements of the Hamiltonian of the system are dependent on $t_4$ and $k$. The system will be chiral symmetric once $t_4 = 0$ (one-dimensional lattice chain). The eigenvalues and eigenvectors of the Hamiltonian can be calculated with these parameters, which read

\begin{align}
	\label{eq:7}
	E_{\pm}(k)&=\pm \sqrt{d_{x}^{2}(k)+d_{y}^{2}(k)}+d_{0}(k),\\
	\label{eq:8}
	\left|\psi_{\pm}(k)\right\rangle&=\left(\pm \frac{\sqrt{d_{x}^{2}(k)+d_{y}^{2}(k)}}{d_{x}(k)+i d_{y}(k)}, 1\right)^{\mathrm{T}}.
\end{align}
\begin{figure}[htb]
	\centering
	\includegraphics[width=\linewidth]{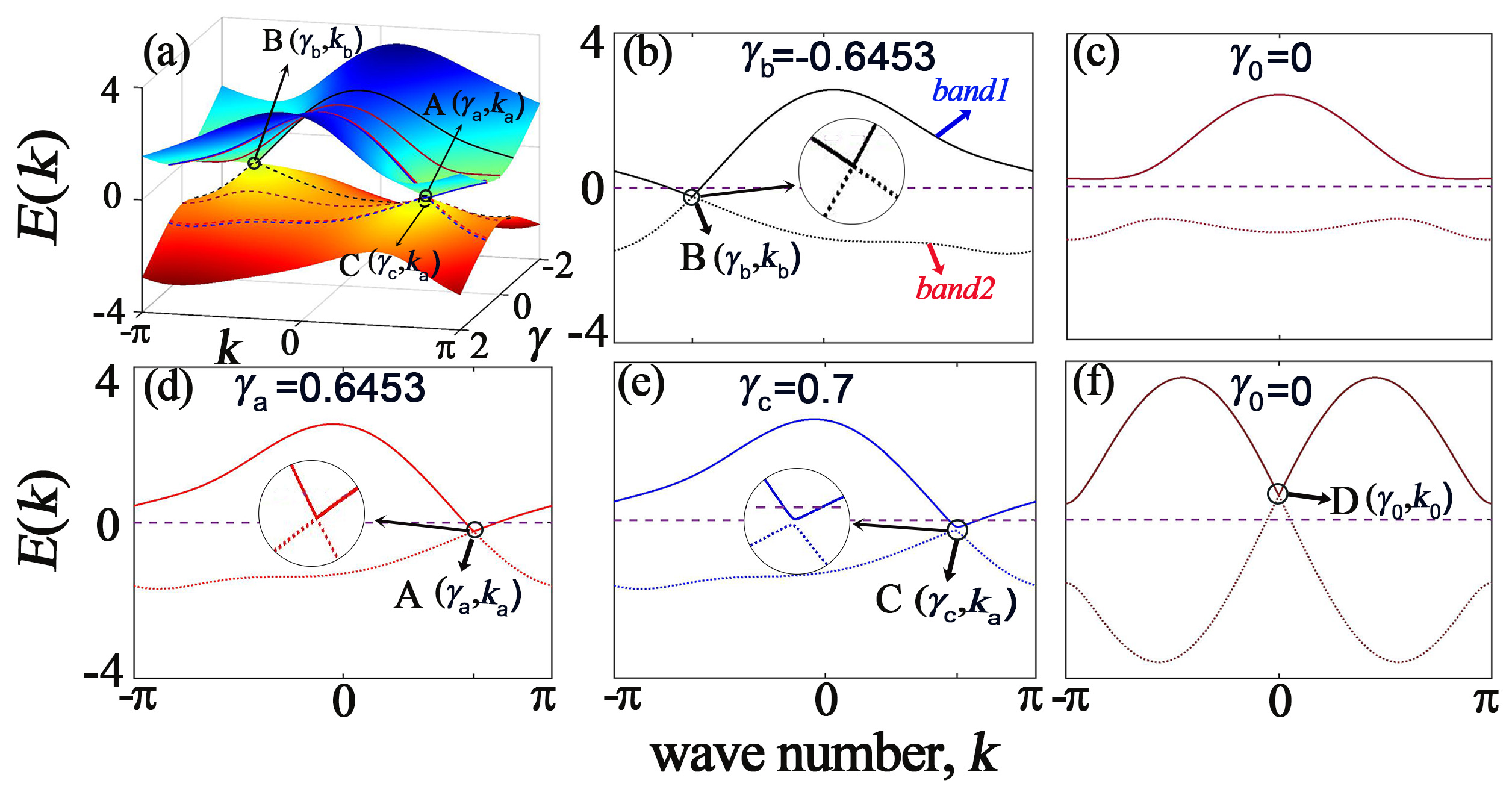}
	\caption{Band structures of the system versus the wavevector $k$ and imaginary coupling coefficient $\gamma$. (a) Curved surface of energy bands, where $k_a=0.626\pi=-k_b$. 
		(b)-(e) The energy band structure under different $\gamma$ with other parameters $t_1=0.5$, $t_2=1$, $t_3=0.3$, $t_4=0.3$. (f) The energy band with one degenerated DP with $t_1=0.5$, $t_2=1.5$, $t_3=-2$, $t_4=0.3$, where $k_0=0$.}
	\label{fig.2}
\end{figure}

The specific band diagram of the bulk Hamiltonian is shown in Fig.~\ref{fig.2}. The wave number $k$ ranges from $-\pi$ to $\pi$, which is exactly one period of the dispersion relation. Therefore, the dispersion relation is sufficient to reflect the variation law of the whole energy band. The energy band structure in the first Brillouin zone with $k$ and $\gamma$ is shown in Fig.~\ref{fig.2}(a). When $\gamma =\gamma_0=0$, the energy band is even symmetrically about $k =k_0= 0$, and the curve changes slightly at $k_0$. Figure.~\ref{fig.2}(b,c,d,e) more intuitively shows the change of the energy band with different $\gamma$. As $|\gamma|$ increases, the energy band is asymmetrical about $k_0$ [see Fig.~\ref{fig.2}(b) and (d)]; the two energy bands gradually intersect to form two DPs both in the left-hand and right-hand sides of $\gamma_0$ and $k_0$, respectively. These two DPs at points A($\gamma_a$, $k_a$) and B($\gamma_b$, $k_b$) is centrosymmetric about the origin ($\gamma_0 $ and $k_0 $), so consistent in nature, where $\gamma_b$= -$\gamma_a$ and $k_b$ = -$k_a$ [see the solid black circles in Fig.~\ref{fig.2}(a)]. This can be proved theoretically from Eq.~(\ref{eq:7}) that $E_{+k}(+\gamma) = E_{-k}(-\gamma)$. Further increasing the value of $|\gamma|$, e.g., $|\gamma| = 0.7$, the band gap gradually enlarged from 0. The closing of the energy bands will inevitably lead to a topological phase transition~\cite{lu2014topological}. As a special case, the degeneracy of these two DPs can finally be achieved when the coupling coefficients satisfy $t_1+t_2+t_3=0$ and $\gamma=0$ [see Fig.~\ref{fig.2}(f)]. At this point (point D), the DPs turn into a high symmetric one~\cite{neto2009electronic}.

The transport behavior of light can be predicted from the energy band structure. When a light wave is excited at an arbitrary waveguide position in the photonic lattice, the transmission direction of the light is along the orthogonal direction of the tangent of the energy band curves. For example, when a light beam is normal incident into the lattice with $\gamma_0$ [Fig.~\ref{fig.2}(c)], the transmission directions of light waves' upper and lower layers are the same. Accordingly, the light wave transmission will be unique due to the particular energy band structures at points A or B.

\section{TOPOLOGICAL DIAGRAM AND ENERGY SPECTRUM}
\begin{figure}[htb]
	\centering
	\includegraphics[width=\linewidth]{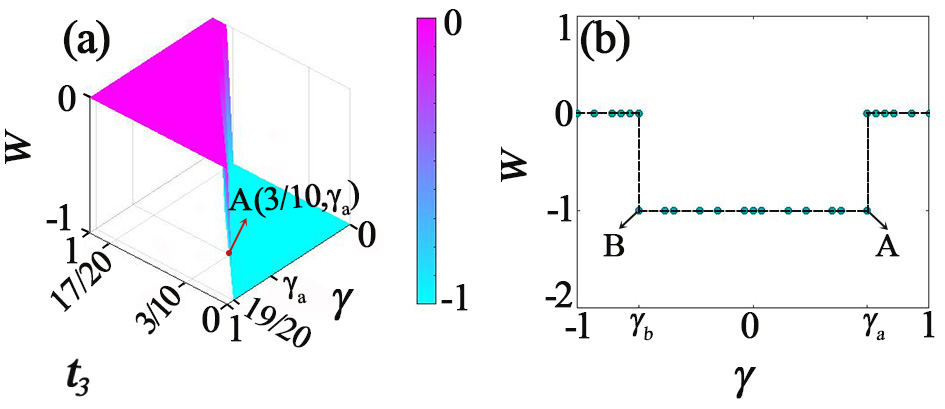}
	\caption{Numerical calculated winding number $W$. (a) Step surface of $W$ versus the interlayer coupling coefficient $t_3$ and imaginary coupling coefficient $\gamma$. The sky blue area ($W=-1$) corresponds to the non-trivial topological state, and the purple area ($W=0$) corresponds to the trivial topological state. (b) Profile of $W$ with $t_3 = 3/10$. Other parameters are $t_1=1/2$, $t_2=1$, $t_4=3/10$.}
\label{fig.3}
\end{figure}
We predict the topological phase transition of the system with topological invariants. In Hermitian systems, the winding number($W$) or Chern number are often calculated as topological invariants to characterize different topologies. To calculate $W$ directly from the body momentum space Hamiltonian, one can see from Eq.~(\ref{eq:8}), noting that the main diagonal elements do not affect the eigenvectors of the system~\cite{Ji2020}. Therefore, $W$~\cite{Ji2020,asboth2016short} of the system is easy to check that
\begin{equation}\label{eq:9}
	W=\frac{1}{2 \pi i} \int_{-\pi}^{\pi} d k \frac{d}{d k} \log h(k) ,
\end{equation}
where $h(k)=d_{x}(k)-i d_{y}(k)$. For the non-reciprocal Hermitian model, $W$ takes -1 when there are topological edge states and 0 when there are not, depending on the interlayer coupling coefficient $t_3$ and imaginary coupling coefficient $\gamma$. The phase diagram is shown in Fig.~\ref{fig.3}(a). It is shown $W = -1$ corresponding to a topological non-trivial state when $t_1=t_2/2,\gamma<19/17 t_3+19/20$ (sky blue area), and $W =0$ corresponding to a topological trivial state when $t_1=t_2/2,\gamma>20/17t_3+17/20$(purple area). The numerical results for $t_3=3/10$ is shown in Fig.~\ref{fig.3}(b). In this case, $A$ and $B$ are the topological phase transition points, namely DPs, consistent with the double DPs A and B in Fig.~\ref{fig.2}(a). This non-zero $W$ in the system indicates that the system is topological non-trivial, supporting topological edge modes of the Hermitian system. 
\begin{figure}[h]
	\centering
	\includegraphics[width=\linewidth]{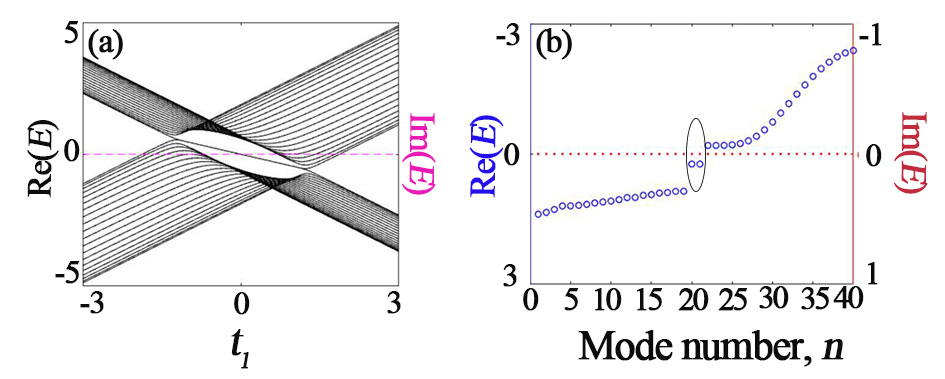}
	\caption{Energy spectrum. (a) Real (left axis Re($E$) in black) and imaginary (right axis Im($E$) in pink) spectra as a function of $t_1$ in the model with open boundary. (b) Real (left axis Re($E$) in blue circles ) and imaginary (right axis Im($E$) in red dots) parts of the eigenvalues of near-zero modes with the non-trivial phase. Here, the total number of the waveguides is $N=40$, $\gamma=0$, $t_1=0.5$ and $t_3=0.3$.}
\label{fig.4}
\end{figure}

We further derived the non-trivial topological states from near-zero modes. The open-boundary energy spectrum is shown in Fig.~\ref{fig.4}(a). The real part of the spectrum is X-shape distributed [see the black curves in Fig.~\ref{fig.4}(a), while the imaginary part is always zero (see the purple dotted line in Fig.~\ref{fig.4}(a)). A pair of near-zero modes locates in the center of the bandgap verifying the existence of topological edge states. Figure.~\ref{fig.4}(b) shows the calculated eigenvalues for the open chain in the non-trivial phase to see near-edge modes as $t_1=0.5$ and $t_3=0.3$. The two modes with $E = -0.2308$ (mode number $N = 20$ and 21, respectively) reside in the band gap, and the imaginary parts of energy are zeroes. Note that the topological condition will be broken once $\gamma$ beyond the range of [$\gamma_b, \gamma_a$], the topological near-zero mode will disappear, and the band gap will reappear and be wider.

\section{TOPOLOGICAL EDGE STATES}
To further investigate the energy distribution for the topological state, the stationary Schr\"{o}dinger-like equations are employed as follows
\begin{align}
	\label{eq:10}
	Ea_n&=(t_1+i\gamma)b_n+t_4(a_{n+1}+a_{n-1})+t_3b_{n+1}+t_2b_{n-1},\\
	\label{eq:11}
	Eb_n&=(t_1-i\gamma)a_n+t_4(b_{n+1}+b_{n-1})+t_2a_{n+1}+t_3a_{n-1}.
\end{align}

Assuming that the amplitude ratio within layers is $\beta=a_{n+1}/a_n=b_{n+1}/b_n$ and between two layers is $\alpha=b_n/a_n$. Thus, one can obtain
\begin{align}
	\label{eq:12}
	E&=(t_1+i\gamma)\alpha+t_4(\beta+\beta^{-1})+t_3\alpha\beta+t_2\alpha\beta^{-1},\\
	\label{eq:13}
	E&=(t_1-i\gamma)\alpha^{-1}+t_4(\beta+\beta^{-1})+t_2\beta\alpha^{-1}+t_3\alpha^{-1}\beta^{-1}.
\end{align}

In the limit of $E\rightarrow 0$, we have the boundary conditions
\begin{align}
	\label{eq:14}
	(t_1+i\gamma)b_0+t_4a_1+t_3b_1&=0,\\
	\label{eq:15}
	(t_1-i\gamma)a_N+t_4b_{N-1}+t_3a_{N-1}&=0.
\end{align}

Substituting Eqs.~(\ref{eq:14}) and (\ref{eq:15}) into Eqs.~(\ref{eq:12}) and (\ref{eq:13}), one can learn that $\beta$, $\alpha$ equals to
\begin{equation}
	\left.\begin{array}{l}
		\beta_{left}=\frac{i\gamma+t_1}{t_2-t_3}, \\
		\alpha_{left}=-\frac{t_4}{t_2},
	\end{array}\right\} \text { left input }
\end{equation}
\begin{equation}
	\left.\begin{array}{l}
	\beta_{right}=\frac{i(t_2-t_3)}{\gamma+it_1}, \text{or}\quad 0,\\
		\alpha_{right}=-\frac{t_2}{t_4}, \text{or} -\frac{t_3}{t_4},
	\end{array}\right\} \text { right input }
\end{equation}
here, $\beta_{left}$ and $\alpha_{left}$, $\beta_{right}$ and $\alpha_{right}$ are the amplitude ratio for the left and right incident cases, respectively.

For the left incidence, it can be deduced that the field distribution satisfies
\begin{align}
	\label{eq:20}
	|a_n|^2&=|\beta_{left}|^{2n}|a_0|^2=(\frac{\gamma^2+t_1^2}{(t_2-t_3)^2})^{n}|a_0|^2,\\
	\label{eq:21}
	|b_n|^2&=|\alpha_{left}|^{2}|\beta_{left}|^{2n}|a_0|^2 =(\frac{t_4}{t_2})^{2}(\frac{\gamma^2+t_1^2}{(t_2-t_3)^2})^{n}|a_0|^2.
\end{align}

For a localized boundary state on the left edge, $|\beta_{left}| <1 $ and $|\alpha_{left}| <1$ always hold. Therefore, Eqs.~(\ref{eq:20}) and ~(\ref{eq:21}) indicate that $a_n$ and $b_n$ both decay exponentially with the cell number $n$. As for a concrete case of $t_2=1$ and $t_4=0.3$, $|b_n|^2$ is nine percent of $|a_n|^2$ showing that most of this energy exists in the waveguides of layer I.

Corresponding to it is the right incident case. It can be deduced that the field distribution satisfies
\begin{align}
	\label{eq:22}
	|a_n|^2&=(\frac{1}{|\alpha_{right}|})^2(\frac{1}{|\beta_{right}|})^{(N/2-n)}|b_N|^2 \notag\\
	&=(\frac{t_4}{t_2})^{2}(\frac{\gamma^2+t_1^2}{(t_2-t_3)^2})^{(N/2-n)}|b_N|^2,\text{or}\notag\\
	&=(\frac{t_4}{t_3})^{2}(\frac{\gamma^2+t_1^2}{(t_2-t_3)^2})^{(N/2-n)}|b_N|^2, \text{$N$ is even}\\
	\label{eq:23}
	|b_n|^2&=(\frac{1}{|\beta_{right}|})^{(N/2-n)}|b_N|^2\notag\\
	&=(\frac{\gamma^2+t_1^2}{(t_2-t_3)^2})^{(N/2-n)}|b_N|^2, \text{$N$ is even}.
\end{align}

One can learn that $|\beta_{right}|>1$ and $|\alpha_{right}|>1$ should always erect for the right-hand side localization on the edge. Therefore, $t_3=t_4=0.3$ is invalid. It can be seen from Eqs.~(\ref{eq:22}) and (\ref{eq:23}) that the energy will be exponentially decreased from the right-hand edge. Different from the left-hand incidence, most energy will be localized on the right-most edge in the waveguides of layer-II. Thus, such a system can achieve a localized topological edge state on the left edge in layer-I for the left incidence and a localized topological edge state on the right edge in layer-II for the right incidence. In other words, such a system can be used to obtain two near-zero modes, which differs from the topological edge modes in Su–Schrieffer–Heeger (SSH)  model~\cite{asboth2016short}. 

\begin{figure}[tp]
	\includegraphics[width=\linewidth]{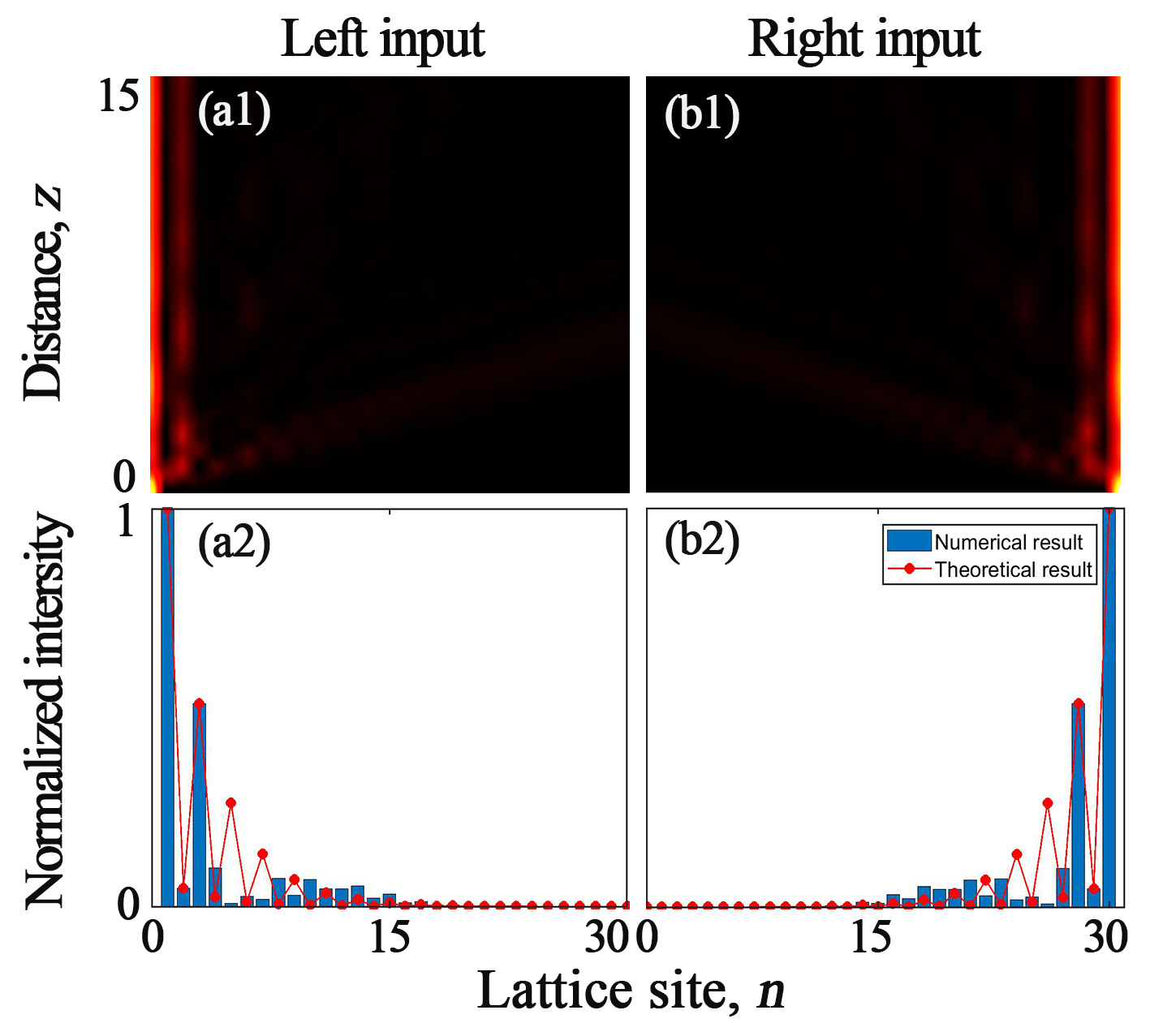}
	\hspace{0.1cm}
	\caption{Numerically calculated light wave evolution and output energy distribution in the system. Here, $\gamma=0, N=30$, $t_1=0.5$, $t_2=1$, $t_3=0.3$, $t_4=0.3$.}
	\label{fig.5}
\end{figure}

To see the characteristics of the system more vividly, we investigate the wave dynamics of Hermitian edge modes in our system, which can be separately stimulated by choosing suitable initial excitation. We launch a Gaussian beam with a function of $a_n(z=0)=A^{'}e^{-(n-n_0)^2/{w}^2+ ik_0n }$, where $A^{'}$ is the magnitude, $n_0$ is the number of incident waveguide, and $w$ is the width of the Gaussian beam. Figures.~\ref{fig.5}(a1,b1) present the wave evolution for different initial conditions as waves are injected from a single site. When waves are injected from the site at the left termination of layer-I, as shown in Fig.~\ref{fig.5}(a1), the edge modes from the Hermitian space are stimulated. We can see some waves are confined at the left edge in layer-I with uniform intensity. After enough evolution distance, the energy becomes stable without dissipation. When waves are injected from the site at the right termination of layer-II, as shown in Fig.~\ref{fig.5}(b1), most of the light energy distributes in the right-most waveguide in layer-II, which is consistent with the aforementioned theoretical predictions. To better understand the properties of our system, the projection of energy distribution at the output facet is calculated both numerically and theoretically [see Figs.~\ref{fig.5}(a2,b2)]. Obviously,
the numerical and theoretical results of the light transmission distribution are in good agreement.
\section{group velocity and wave dynamics}
\begin{figure*}[htp]
	\includegraphics[width=\linewidth]{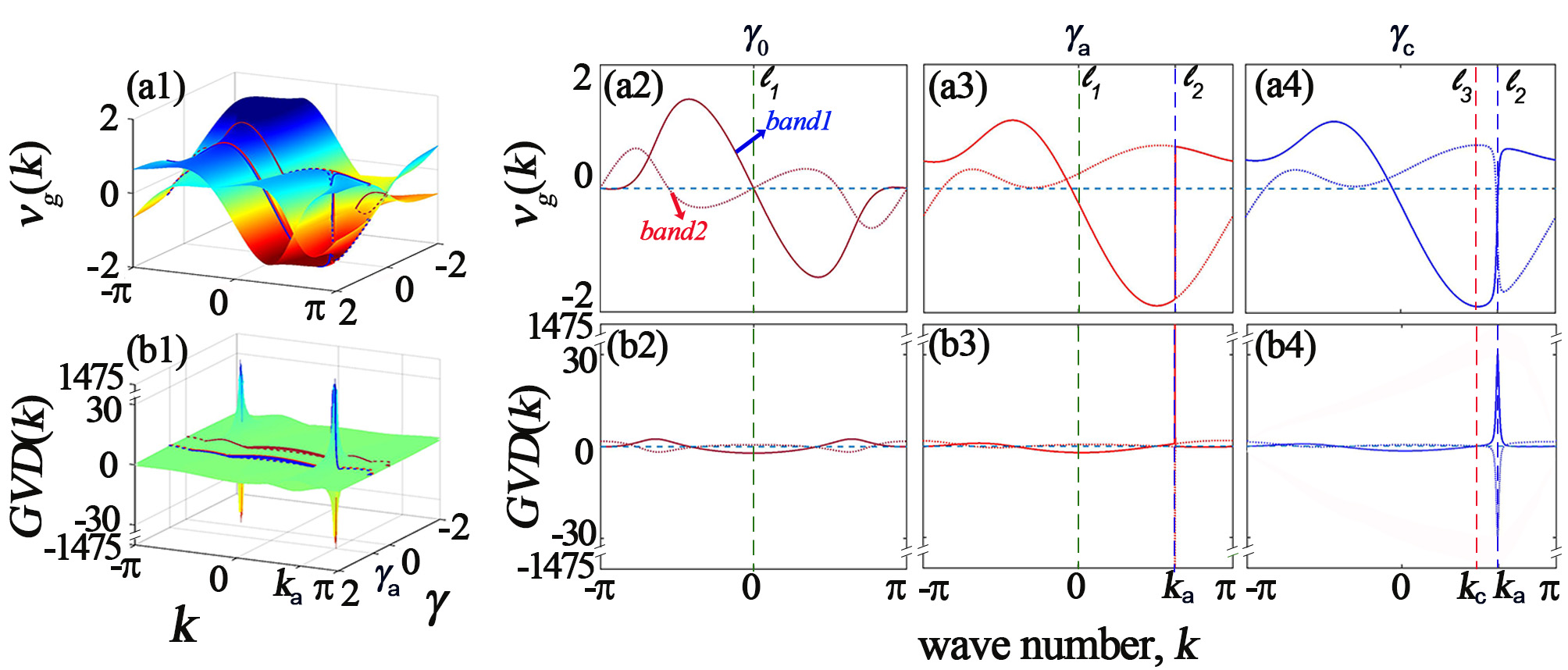}
	\hspace{0.1cm}
	\caption{Distributions of group velocity and group velocity dispersion versus the wavevector $k$ and imaginary coupling coefficient $\gamma$. (a2, b2) $\gamma_0 =0$. (a3, b3) $\gamma_a = 0.6453$. (a4, b4) $\gamma_c = 0.7$. $l_1$, $l_2$ and $l_3$ correspond to $k_0=0$, $k_a=0.626\pi$ and $k_c=\pi /2$, respectively. Other parameters are $t_1=0.5$, $t_2=1$, $t_3=0.3$, $t_4=0.3$.}
	\label{fig.6}
\end{figure*}
\begin{figure*} [htp]
	\begin{align}
		\label{eq:24}
		v_{g_{\pm}}&=\frac{\partial E_{\pm}}{\partial k}=-2 t_4 \sin k\pm\frac{2 \gamma(-t_2+t_3)\cos k-2t_1(t_2+t_3)\sin k-4t_2t_3 \sin(2k)}{2 \sqrt{\gamma^{2}+t_1^{2}+t_2^{2}+t_3^{2}+2t_1(t_2+t_3) \cos k+2t_2t_3\cos(2k)+2 \gamma(-t_2+t_3)\sin k}}=v_{g_1} \pm v_{g_2}.\\
		\label{eq:25} 
		GVD_{\pm}&=\frac{\partial E_{\pm}^2}{\partial^2 k}=-2 t_4 \cos k\pm\frac{-2 t_1(t_2+t_3)\cos k-8t_2t_3\cos(2k)-2 \gamma(-t_2+t_3)\sin k }{2 \sqrt{\gamma^{2}+t_1^{2}+t_2^{2}+t_3^{2}+2t_1(t_2+t_3) \cos k+2t_2t_3\cos(2k)+2 \gamma(-t_2+t_3)\sin k}} \notag\\
		&\mp\frac{[2 \gamma(-t_2+t_3)\cos k-2t_1(t_2+t_3)\sin k-4t_2t_3 \sin(2k)]^{2}}{4[\gamma^{2}+t_1^{2}+t_2^{2}+t_3^{2}+2t_1(t_2+t_3) \cos k+2t_2t_3\cos(2k)+2 \gamma(-t_2+t_3)\sin k]^{3 / 2}}.
	\end{align}
\end{figure*} 

Further study of light wave dynamics in Hermitian photonic lattices is carried out by analytically solving the group velocity ($v_g$). Eq.~(\ref{eq:24}) indicates that $v_g$ has two different values ​​under the same incident wave vector, but the $v_g$ is the same when $v_{g_2} = 0$. The surface distribution of $v_g$ with the wave vector $k$ and $\gamma$ is shown in Fig.~\ref{fig.6}(a1). The three curves with $\gamma$ from -2 to 2 in Fig.~\ref{fig.6}(a1) correspond to Figs.~\ref{fig.6}(a2)-(a4) on the right page. When it is a reciprocal system ($\gamma = \gamma_0 =0$), $v_g$ for the two bands in the first Brillouin zone are oddly symmetric about the origin of ($k_0$, $\gamma_0$). Clearly, $v_g$ for the second band changes more rapidly than for the first band. Besides, fast light, slow light, and light stops can all be achieved in one cycle [see Fig.~\ref{fig.6}(a2)]. In particular, two $v_g$ profile change suddenly at DP [point A in Figs.~\ref{fig.2}(a) and (d), Fig.~\ref{fig.3}(b)] when $\gamma=\gamma_a$ as shown in Fig.~\ref{fig.6}(a3).  When $\gamma$ is increased to $\gamma_c$, the $v_g$ profiles change gradually at DP [see the dashed line $l_2$ in blue for $k_a$ in Fig.~\ref{fig.6}(a4)], corresponding to the point C in Fig.~\ref{fig.3}(e).

On this basis, we obtain the group velocity dispersion ($GVD$), as shown in Eq.~(\ref{eq:25}). Obviously,  $GVD$ is related to the imaginary coupling coefficient $\gamma$ and the wave vector $k$. The surface picture is given in Fig.~\ref{fig.6}(b1). Two particular peaks on the surface correspond exactly to points A and B in Fig.~\ref{fig.2}(a). Figs.~\ref{fig.6}(b2) to (b4) show the profiles of $GVD$ versus the wavevector $k$ for three $\gamma$.  When $\gamma=\gamma_0$, the interlayer coupling is reciprocal, which gives an evenly symmetric distribution about the origin $k=0$ with a dispersion of almost zero [Fig.~\ref{fig.6}(b2)]. Further increasing the value of $\gamma$ to $\gamma_a$, very large fluctuations appear at $k_a$ for both of the two bands [Fig.~\ref{fig.6}(b3)]. At this mutation point A, the values of these two $GVD$ are tremendous (close to 1,475) but with opposite signs [see the dashed blue line in Fig.~\ref{fig.6}(b3)]. As a result, this will give an unstable propagation of light waves under these incident conditions and may lead to beam splitting in waveguides of both layers. The magnitudes of the $GVD$ peaks contract to less than $\pm 30$ and $GVD=0$ holds for both the two bands simultaneously at $ k_c= \pi/2$ when $\gamma=\gamma_c$ [see Fig.~\ref{fig.6}(b4)],  which means that the light beam may experience diffractionless propagation when the light is injected with wave vector $k_c$.

\begin{figure}[tb]
	\centering
	\includegraphics[width=\linewidth]{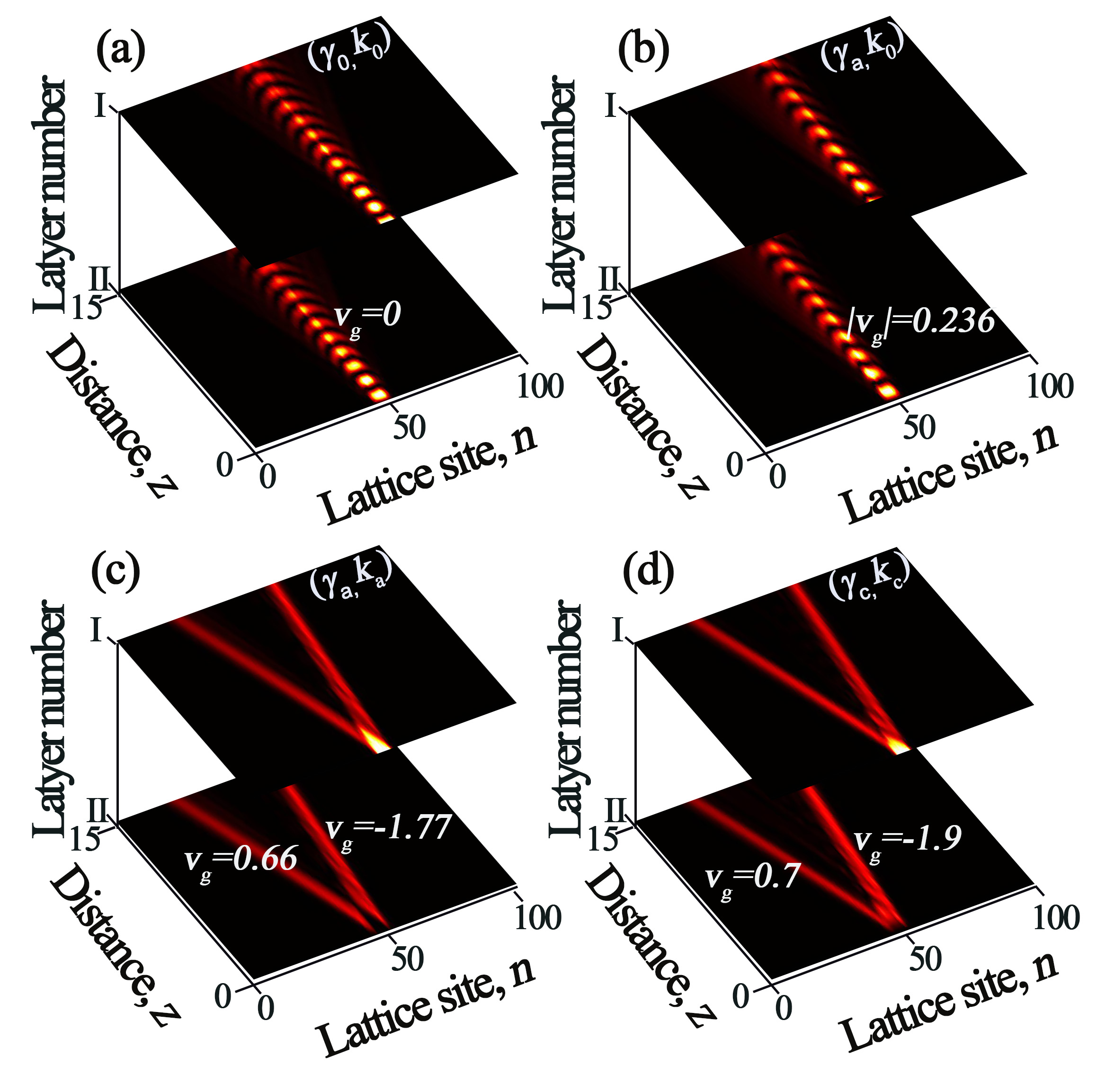}
\caption{Wave dynamics for non-reciprocal Hermitian system. The width of Gaussian beam is $\omega = 3$. Other parameters are the same as that used in Fig.~\ref{fig.6}.}
\label{fig.7}
\end{figure}

We now study the light transmission behavior numerically. As shown in Figs.~\ref{fig.7}(a) and (b), a Gaussian wave packet is, respectively, centered initially at the waveguide site $N_a=50$ in layer I with $k_0=0$. Light transmits forward and oscillates in the waveguides between two layers because of the interlayer coupling effect. However, slight contraction in the transverse direction exists during the propagation, and the calculated $v_g$ also from 0 to $\pm 0.236$ due to the action of the non-reciprocal coupling effect arising from $\gamma$ [see the green dashed line $l_1$ in Figs.~\ref{fig.6}(a2) and (a3)]. Therefore, the optical wave oscillation behavior between two layers is not changed dramatically, but the $v_g$ may be tuned by introducing the imaginary coupling coefficient for the same incident angle. When the Gaussian beam is injected into the waveguides in layer I under conditions of $k = k_a$ and $\gamma = \gamma_a$ [Fig.~\ref{fig.7}(c)], light splits both in layer I and in layer II due to degeneracy of the energy band and the abrupt change in the $v_g$ at DP [Fig.~\ref{fig.2}(d) and Fig.~\ref{fig.6}(b3)]. Besides, all these four lobes are barely broadened because of their tiny dispersion around DP [see the curves around the dashed blue line in Fig.~\ref{fig.6}(b3)]. Further calculation shows that the $v_g$ for the left and right lobes are different, $v_g=0.66$ for the left side and $v_g=-1.77$ for the right side, showing that such a configuration may achieve beam bifurcation with slow and fast light diffractionfree transmission. As a comparison, the propagation of a Gaussian beam under the conditions of $k_c$ and $\gamma_c$ is also investigated [Fig.~\ref{fig.7}(d)]. It is found that the light wave is bifurcated without diffraction because of the zero dispersion and the coupling effects, which is consistent with the previous predictions. The $v_g$ for the left and right bifurcated lights are $v_g=0.7$ and $v_g=-1.9$, respectively, indicating the mutation point of the $GVD$ does not mean the maximum point of the $v_g$. In other words, we have realized two mechanisms to achieve bifurcated diffractionfree light in one system by tuning the imaginary coupling coefficient $\gamma$ and the wavevector $k$.

\section{Conclusions}
In conclusion, we have theoretically and numerically demonstrated that a two-layered photonic lattice with non-reciprocal coupling in each unit cell could sustain a real energy spectrum. In such a system, a pair of Dirac points corresponding to topological phase transition points can survive and be adjusted to a single one by tuning the coupling coefficient. Furthermore, our system supports topological edge states existing on the left or right boundaries of respective lattice layers. In addition, we show the ways to control the group velocity and group velocity dispersion by the imaginary coupling and wavevector. Two types of bifurcated diffractionfree beams with tunable group velocity are achieved by employing the properties at the Dirac point and the non-reciprocal coupling effects. The proposed imaginary coupling-induced group velocity regulation mechanism enriches the physics of group velocity manipulation and may have potential applications in designing novel types of group velocity delay compensators and photonic directional couplers. 

\section{Acknowledgement}
We gratefully acknowledge financial support by the National Natural Science Foundation of China (NSFC)(N0.1217040857).

\nocite{*}

\bibliography{ref}

\end{document}